\documentclass[12pt]{article}
\newcommand{\e}{\mathrm{e}}
\newcommand{\be}{\begin{equation}}
\newcommand{\ee}{\end{equation}}
\newcommand{\bea}{\begin{eqnarray}}
\newcommand{\eea}{\end{eqnarray}}

\newcommand{\Tr}{\mathop{\mathrm{Tr}}\nolimits}
\newcommand{\dd}{\mathrm{d}}
\newcommand{\ket}[1]{\left|#1\right\rangle}
\newcommand{\bra}[1]{\left\langle #1\right|}

\newcommand{\unit}{1\!\!1}
\newcommand{\NN}{I\!\!N}

\begin{document}

\title{Discrete Nonlocal Waves}

\author{Ciprian Acatrinei\\
        Department of Theoretical Physics \\
        Horia Hulubei National Institute for Nuclear Physics\\
        Bucharest, Romania} 

\date{7 November 2012}

\maketitle

\begin{abstract}

We study generic 
waves without rotational symmetry
in (2+1) - dimensional noncommutative scalar field theory.
In the representation chosen, the radial coordinate is naturally rendered discrete.
Nonlocality 
along this coordinate, 
induced by noncommutativity,
accounts for the angular dependence of the fields. 
The exact form of standing and propagating waves on such a discrete space 
is found in terms of finite series. 
A precise correspondence is established between the degree of nonlocality
and the angular momentum of a field configuration.  
At small distance no classical singularities appear, even at the location of the sources.
At large radius one recovers the usual commutative behaviour.

\end{abstract}

\section{Introduction and Overview}

Over the years, various attempts were made to introduce field theories which are nonlocal  
over a continuous or a discrete space. The hopes related to such a program were rarely fulfilled,
often due to technical difficulties or structural issues.

A relatively recent 
example of a nonlocal field theory (FT) without
obvious inconsistencies and which can be handled analytically is given by 
noncommutative (NC) field theory \cite{ncft}. 
One assumes that fields live over a space which is not commutative; 
for a plane one has $[x,y]=i\theta \neq 0$.
In the most frequent approach to NC FT, nonlocality is simply attributed to 
the infinite number of derivatives in the star-product induced by the NC algebra.

Using operatorial methods however, 
nonlocality can be quantified in a more precise way.
Namely, the harmonic oscillator 
basis is used to naturally discretize the radial coordinate. 
Then, NC is traded for nonlocality along this discretized coordinate.
Finally, precisely this nonlocality turns out to account for the angular (nonradial)
dependence of the fields.
Our goal is to understand free radial and nonradial, local and nonlocal, 
oscillation and propagation phenomena
in such a framework, 
with discreteness and nonlocality explicit at the level of the degrees of freedom 
but manageable analytically.

In the regime exactly opposite to the one we consider, namely when the kinetic term is neglected, 
interesting approximate radially symmetric solutions were obtained for scalar NC FT \cite{gms}.
Subsequent perturbation theory in $\frac{1}{\theta}$ 
confirmed \cite{djn}-\cite{sol} their existence but  did not take the kinetic term 
on NC space fully into account.
In spite of extensive work \cite{gms}-\cite{A1},
including exact results obtained for nonscalar theories \cite{pol}-\cite{lp},
no complete account of free propagation on a discrete space seems available.
In particular, one needs explicit solutions without radial symmetry 
to include configurations with nonzero angular momentum.

Our contribution is the following. We provide new exact finite series expressions
for radially and nonradially symmetric oscillations and propagating waves,
include sources and display the commutative limit.
Along these technical results two significant conclusions arise.
First, no classical divergences appear at small distance; 
fields stay finite even at the location of the sources. 
Second, field configurations turn out to be local when they have zero angular momentum 
and nonlocal whenever their angular momentum is nonzero. Actually the angular momentum and the 
(appropriately defined) degree of nonlocality of a field configuration can be identified 
up to a multiplicative constant. This last statement is proved through an extension
of Noethers's theorem to operatorially valued field configurations.

The paper is organized as follows. 
Section 2 introduces scalar NC FT, 
the radial basis which discretizes the field equations of motion
and the simple procedure which trades NC for nonlocality.
The nature of planar NC angular momentum is clarified in Section 3 
through a Noether-type argument for operatorially valued-fields.
A precise connection is established between the degree of nonlocality 
and the angular momentum of a field configuration.
In Section 4 exact solutions are derived for the discrete field equations in terms of finite series.
This is the main technical part of the paper.
In Section 5 sources are included - without singularities appearing. 
The commutative limit is taken in Section 6, clarifying the nature of the NC solutions
derived previously and confirming the precise way in which 
nonlocality is related to nonvanishing angular momentum.
The Appendix collects known and less-known formulae required for the technical proofs.

The general notation used is quite standard: $n$, $k$ represent non-negative integers,
the set of which is occasionally denoted by $\NN$;
the gamma function obeys $\Gamma(z+1)=z\Gamma(z)$;
the binomial coefficient is $C_n^k=\frac{n!}{(n-k)!k!}$.
A discrete version of the logarithm, 
$H_n=1+1/2+\dots+1/n$,
plays an important role in what follows. $\delta_{n,k}\equiv\delta(n,k)$ is the Kronecker symbol,
which is one when $n=k$ and zero otherwise.
For a local field $\phi$, $\phi(n)\equiv\phi_n$ denotes its value at the point $n$.
For a bilocal field $\Phi$ that depends on {\em two} points, $\Phi(n,n')\equiv\Phi_{n,n'}$ 
denotes its value when the first point is taken at $n$ and the second at $n'$.

\section{Equations of motion}

Consider a (2+1)-dimensional
scalar field $\phi$, depending on space coordinates forming
a Heisenberg algebra (time is commutative and remains continuous):

\be
\phi(t,\hat{x},\hat{y}), \qquad [\hat{x},\hat{y}]=i\theta\hat{\unit} \; . \label{cr}
\ee$\theta$ is a constant with the dimensionality of an area.
The scalar field $\phi$ is consequently a time-dependent operator 
acting on the Hilbert space ${\cal H}$ on which the algebra is represented. 
Since (\ref{cr}) implies
$$
[\hat{x}, \phi(\hat{x},\hat{y})]=i\theta\frac{\partial \phi}{\partial \hat{y}},
\qquad \qquad
[\hat{y}, \phi(\hat{x},\hat{y})]=-i\theta\frac{\partial \phi}{\partial \hat{x}},
$$
the field action, written in operatorial form, reads
\be
S=\int\dd t \Tr_{{\cal H}} 
\left (
\frac{1}{2}\dot{\phi}^{\dagger}\dot{\phi}
+\frac{1}{2}[\hat{x}, \phi^{\dagger}][\hat{x}, \phi] 
+\frac{1}{2}[\hat{x}, \phi^{\dagger}][\hat{y}, \phi]
+V(\phi^{\dagger}\phi)
\right )
\label{action}
\ee
allowing for classical solutions $\phi$ which are non-Hermitian, $\phi^{\dagger}\neq \phi$.
We take $V(\phi^{\dagger}\phi)=0$ here. 
The equations of motion for the field $\phi$ are then
\be
\ddot{\phi}+
\frac{1}{\theta^2}[\hat{x},[\hat{x},\phi]]+\frac{1}{\theta^2}[\hat{y},[\hat{y},\phi]]=0. 
\label{em}
\ee

In Cartesian coordinates, the solutions are plane waves
\be
\phi \sim e^{i(k_x \hat{x} + k_y \hat{y})-i\omega t}, \quad k_x^2+k_y^2=\omega^2,  
\label{plane}
\ee
formally identical to the commutative ones. 
In fact, due to the operators in the exponent, such waves have bilocal character \cite{A2}, 
in agreement with the considerations to follow.
If a mass term $m^2\phi^{\dagger}\phi/2$ is inserted in Eq. (\ref{action}), 
$\omega^2$ should be replaced by $\omega^2-m^2$.

If the physical situation requires polar coordinates 
(a source emiting radiation, a circular membrane oscillating),
then the harmonic oscillator, or radial, basis $\{\ket{n}\}$ 
\be
\hat{N}\ket{n}=n\ket{n}, \quad \hat{N}=\hat{a}^{\dagger}\hat{a}, \quad
\hat{a}=\frac{1}{\sqrt{2\theta}}(\hat{x}+i\hat{y}), 
\quad n=0,1,2,\dots
\label{basis}
\ee 
is the natural one, cf. e.g. \cite{AS}, and the equations of motion become
\be
\ddot{\phi}+\frac{2}{\theta}[\hat{a},[\hat{a}^{\dagger},\phi]]=0. \label{em2}
\ee
$\hat{N}=\frac{1}{2}(\frac{\hat{x}^2+\hat{y}^2}{\theta}-1)$ 
is basically (half) the radius square operator, in units of $\theta$.
Its eigenvalues $n$ in (\ref{basis}) correspond to discrete points
with radius growing like $\sqrt{n}$ ($n\sim\frac{r^2}{2\theta}$) for large $n$.
The NC plane is realized via (\ref{basis}) as the semi-infinite discrete space 
of the points labeled by $n$.

{\bf Origin of the recurrence relation}

We can sandwich any equation containing the operatorial field $\phi(\hat{x},\hat{y},t)$
between $\bra{n'}$ and $\ket{n}$ states, eliminating NC in this way. 
The resulting field $\bra{n'}\phi(t)\ket{n}\equiv \phi_{n',n}(t)$ is indeed a commutative object
but also a bilocal one, 
as it generically depends on two different points, $n$ and $n'$.
Consider an harmonic time dependence 
$\phi_{n',n}(t) = e^{i\omega t}\phi_{n',n} $
and use
$\hat{a}\ket{n}=\sqrt{n}\ket{n-1}$, $\hat{a}^{\dagger}\ket{n}=\sqrt{n+1}\ket{n+1}$. 
Eq. (\ref{em2}) implies that the equation of motion for $ \phi_{n',n}$ is
\be
\sqrt{n'+1}\sqrt{n+1}\phi_{n'+1,n+1} +\sqrt{n'}\sqrt{n} \phi_{n'-1,n-1}
-(n'+1+n-\lambda)\phi_{n',n}=0.
\label{difference_eq}
\ee
Above, $\lambda=\frac{\theta}{2} \omega^2$. 
Eq. (\ref{difference_eq}) is a recurrence relation of order two, 
describing the radial classical dynamics of a field which lives on a discrete space. 
The initial angular dependence of 
$\phi(\hat{x},\hat{y},t)$,
due to its dependence on both $ \hat{x}$ and $ \hat{y}$, is not lost. 
It is encoded in the two-index structure of $ \phi_{n',n}$,
as will be seen explicitely in what follows.
The operator $\phi$ is reconstructed from the c-numbers $\phi_{n',n}$ via
\be
\phi=\sum_{n',n\in \NN}\ket{n'}\phi_{n',n}(t)\bra{n}.
\label{opform}
\ee
The bilocality of the fields $\phi_{n',n}$ thus implies the nonlocality of $\phi$.

{\bf Radial symmetry}

If however the field $\phi$ depends only on 
the combination of $\hat{x}$ and $\hat{y}$ given by $\hat{N}$, 
$\phi =\phi (\hat{N})$, we have radial symmetry.  
In this case $\phi$ is diagonal in the $\ket{n}$ basis, 
$\bra{n'}\phi \ket{n}=\phi_{n',n}\delta_{n',n}\;$, and we have a local field.
Defining $\phi_{n,n}\equiv \phi_n $ for simplicity, Eq. (\ref{difference_eq}) implies
\be
(n+1)\phi_{n+1}+n\phi_{n-1}+(\lambda-2n-1)\phi_n=0, \quad n=0,1,2,\dots  \label{em3}
\ee
again with $\lambda=\omega^2\theta /2$. 
The expectation value  $\phi_n\equiv  \bra{n}\phi \ket{n}$ characterizes 
$\phi=\sum_{n=0}^{\infty}\ket{n}\phi_n\bra{n}$ uniformly at radius squared $n$. 
 No angular dependence appears anymore. 
If a single value $\phi_{n_0}$ is nonzero, 
then $\ket{n_0}\phi_{n_0}\bra{n_0}$ describes a field located at $n_0$.

{\bf A more suggestive form}

Returning to the general case (\ref{opform}), consider the situation in which
the first index of $\phi_{n',n}$ is greater than the second, $n'\geq n$.
Define $\phi^{(m)}_{n}\equiv\phi_{n',n}$, with $m\equiv n'-n\geq 0$.  
The classical equation of motion of $\phi^{(m)}_{n}$ follows from (\ref{difference_eq}), 
\be
\sqrt{n+m+1}  \sqrt{n+1}  \phi^{(m)}_{n+1}+ \sqrt{n+m} \sqrt{n} \phi^{(m)}_{n-1}  
+ (\lambda-2n-m-1) \phi^{(m)}_{n}=0.
\label{Phinm}
\ee
The index $m$ stays constant throughout the above second-order difference equation in $n$.
To include the situation in which the second index is greater than the first,
introduce $\phi^{(-m)}_{n}\equiv\phi_{n,n'}$, again with $m\equiv n'-n \geq0$.
$\phi^{(-m)}_{n}$ is easily shown to obey the same equation as $\phi^{(m)}_{n}$.
Consequently, the independent solutions of Eq. (\ref{Phinm}) are sufficient
to characterize completely both  $\phi^{(m)}_{n}$ and $\phi^{(-m)}_{n}$,
provided their boundary conditions are assigned independently. 
To summarize, if the notation
\be
\phi_{n_1,n_2}\equiv {\phi}_{\mbox{\scriptsize min}\{n_1,n_2\}}^{(n_1-n_2)}, 
\qquad m\equiv |n_1-n_2|, 
\qquad n_1,n_2\in \NN
\ee
is adopted, the expression (\ref{opform}) turns into
\be
\phi=\sum_{m=0}^{\infty}a_m\sum_{n=0}^{\infty}\ket{n+m}e^{i\omega t}\phi_n^{(m)}\bra{n}
    +\sum_{m=0}^{\infty}b_m\sum_{n=0}^{\infty}\ket{n}e^{i\omega t}\phi_n^{(-m)}\bra{n+m}.
\label{opformmm}
\ee
Eq. (\ref{Phinm}) implies that configurations with different $m$
can be freely superposed in (\ref{opformmm}), 
as underlined by the insertion of the coefficients $a_m$ and $b_m$, 
determined solely through initial/boundary conditions.
In contrast to $\ket{n}\phi_n^{(0)}\bra{n}$ 
which associates a value $\phi_n^{(0)}$ to the point $n$,
$\ket{n+m}\phi_n^{(m)}\bra{n}$  associates a value $\phi_n^{(m)}$ 
to the {\em two} points $n+m$ and $n$.
Thus, $m$ is a measure of the delocalization of the field configuration it characterizes.
Further statements can be made about the $m$-expansion (\ref{opformmm}) 
without solving the classical equation of motion (\ref{Phinm}), 
as we show next.

\section{Angular Momentum versus Nonlocality}

Ref. \cite{A2} showed that the Cartesian coordinates plane wave excitations (\ref{plane})
of a planar NC FT relate to 
one-dimensional dipoles described by a position $x$ and an extension $\delta x=\theta p_y$,
with $p_y$ the linear momentum in the $y$ direction. 
One cannot use $y$ as a second independent variable due to $[\hat{x},\hat{y}]\neq 0$; 
the conjugate variable $p_y$ turns out to be the natural substitute.

A similar picture can be developed for radial coordinates.
The bilocal quantity $\phi_n^{(m)}\equiv\phi_{n',n}$ 
can be described by two variables: 
a discrete radius squared given by $n$ (by $2n+1$ actually)
and an 'extension' $m\equiv|n'-n|$. The analogy
with the plane wave scenario suggests that $m$ is related to the quantity conjugated to the angle,
i.e. to the planar angular momentum.

To prove this, we adapt Noether's theorem to the operatorial set-up 
of $\phi(\hat{x},\hat{y})$
and obtain the expression for angular momentum in the $x-y$ plane, 
$J_z\equiv J$.
First we identify the generator of rotations in the NC plane. Using
\be
e^{i\alpha \hat{R}}\hat{O}e^{-i\alpha \hat{R}}=
O+i\alpha[R,O]-\frac{\alpha^2}{2!}[R,[R,O]]-i\frac{\alpha^3}{3!}[R,[R,[R,O]]]+\cdots
\ee
and recalling $\hat{N}=\frac{1}{2}(\frac{\hat{x}^2+\hat{y}^2}{\theta}-1)=\hat{a}^{\dagger}\hat{a}$,
we see that 
\be
e^{i\alpha \hat{N}}\hat{x}e^{-i\alpha \hat{N}}=+\hat{x}\cos\alpha+\hat{y}\sin\alpha, \qquad
e^{i\alpha \hat{N}}\hat{y}e^{-i\alpha \hat{N}}=-\hat{x}\sin\alpha+\hat{y}\cos\alpha.
\ee
Consequently, $\hat{N}$ generates rotations in the $x-y$ plane. 
The variation of the field $\phi$ under an infinitesimal rotation is then
\be
\delta \phi \; \equiv \;
e^{i\alpha \hat{N}}\phi(\hat{x},\hat{y})e^{-i\alpha \hat{N}}-\phi(\hat{x},\hat{y})
\quad \simeq \; \; i\alpha[\hat{N},\phi], \;\; \mbox{if} \;\; \alpha\rightarrow 0.
\ee
Due to the trace over the Hilbert space appearing in (\ref{action}),
the field action remains invariant under such unitary transformations.
One can therefore proceed and adapt the usual way of thinking of the Noether theorem 
to this (still classical although) operatorial set-up. 
The conserved charge associated to the invariance under rotations, the angular momemtum,
turns out to be
\be
J_z=Tr_H \; i\dot{\phi}^{\dagger}[\hat{a}^{\dagger}\hat{a},\phi].
\label{J}
\ee
The commutator appearing in (\ref{J}) is particularly simple
in two instances,
\be
\left [\hat{a}^{\dagger}\hat{a},\ket{n+m}\bra{n}\right ]=m\; \ket{n+m}\bra{n}
\ee
and
\be
\left [\hat{a}^{\dagger}\hat{a},\ket{n}\bra{n+m}\right ]=-m\; \ket{n}\bra{n+m}.
\ee
These two equations already show that states 'delocalized' by an amount $m$
are expected to carry $m$ units of angular momentum.
Using them 
and denoting the angular momentum of a field configuration $\phi$ by $J_z\left [\phi\right ]$
we obtain for the expressions entering (\ref{opformmm})
\be
J_z
\left [
\sum_{n\in\NN}\ket{n+m}\phi_n^{(m)}\bra{n}
\right ]
=+m\omega  \; \sum_{n=0}^{\infty}[\phi_n^{(m)}]^2;
\label{j=+m}
\ee
\be
J_z
\left [
\sum_{n\in\NN}\ket{n}\phi_n^{(-m)}\bra{n+m}
\right ]
=-m\omega  \; \sum_{n=0}^{\infty}[\phi_n^{(-m)}]^2.
\label{j=-m}
\ee
Dividing by the  normalization factor $N^{+}_m=\sum_{n=0}^{\infty}[\phi_n^{(m)}]^2$,
respectively by $N^{-}_m=\sum_{n=0}^{\infty}[\phi_n^{(-m)}]^2$,
leaves us with the results $(+m\omega)$, respectively $(-m\omega)$.
No matter what the precise values of $\phi_n^{(m)}$ and $\phi_n^{(-m)} $ are,
the index $m$ determines the angular momentum, which is related exclusively to the
degree of delocalization of a field configuration.
The fact that $N^{+}_m$ and $N^{-}_m$ may be formally infinite 
  [similarly to the situation of commutative 2D radial waves, or even of harmonic 1D waves] 
is irrelevant, since the normalization factors cancel out in the final result.  
If $N^{+}_m$ and $N^{-}_m$ are finite or properly regularized
  [e.g. through an upper cut-off $n_{max}$, 
   or a multiplicative factor $r^{n/2}<1$ suppressing $\phi_n^{(m)}$ at large $n$,
   to be removed only at the end of the calculations] 
we can further define
\be
\phi^{(+m)}\equiv\frac{1}{\sqrt{N^{+}_m}}\sum_{n\in\NN}\ket{n+m}\phi_n^{(m)}\bra{n}, \quad
\phi^{(-m)}\equiv\frac{1}{\sqrt{N^{-}_m}}\sum_{n\in\NN}\ket{n}\phi_n^{(-m)}\bra{n+m}
\label{clar!?}
\ee 
and suggestively write the general solution (\ref{opformmm}) as an $m$-expansion
\be
\phi=\sum_{m\in\NN}[a_m\phi^{(+m)}+b_m\phi^{(-m)}].
\label{gensol}
\ee
The operatorial solution $\phi^{(+m)}$ has delocalization $m$, cf. (\ref{clar!?}),
and angular momentum $+m\omega$, cf. (\ref{j=+m}). Similarly,
$\phi^{(-m)}$ has delocalization $-m$, cf. (\ref{clar!?}),
and angular momentum $-m\omega$, cf. (\ref{j=-m}). 
In consequence, (\ref{gensol}) admits two equivalent interpretations.
From the point of view of the theory defined on the discrete set of points $n$,
it is an expansion in  field configurations with well-defined nonlocality,
more precisely bilocality, $m$. 
From the point of view of planar NC FT, Eq. (\ref{gensol}) takes into
account the nonradial dependence of $\phi$ through an angular momentum expansion.

\section{Bilocal waves via finite series}

In this section we solve Eq. (\ref{Phinm}) 
to obtain the exact form of $\phi^{(m)}_{n}$ and $\phi^{(-m)}_{n}$.
The difference equation (\ref{Phinm}) describes travelling or standing waves 
on the semi-infinite discrete space of points $n\in \NN$.
It is convenient to parametrize its two independent solutions as follows:
\be
\phi^{1(m)}_{n}=\sqrt{\frac{(n+m)!}{n!}}f_{1}(\lambda)u^{(m)}_{n},
\qquad
\phi^{2(m)}_{n}=\sqrt{\frac{(n+m)!}{n!}}f_{2}(\lambda)v^{(m)}_{n}.
\label{param}
\ee
The functions $f_{1}(\lambda)$ and $f_{2}(\lambda)$,
important for normalization, will be given later. 
They can be temporarely neglected since they do not depend on $n$. 
$u^{(m)}_{n}$ and $v^{(m)}_{n}$, which will turn out to be two polynomials in $\lambda$,
are denoted collectively by $\tilde{\phi}_n^{(m)}$.
In consequence, (\ref{param}) amounts for the time being to the substitution
\be
\phi_n^{(m)} = \sqrt{\frac{(n+m)!}{n!}}\tilde{\phi}_n^{(m)}.
\label{subst}
\ee
The field $\tilde{\phi}_n^{(m)}$ then satisfies the simpler recurrence
\be
(n+m+1)\tilde{\phi}_{n+1}^{(m)}+n\tilde{\phi}_{n-1}^{(m)}+(\lambda-2n-m-1)\tilde{\phi}_n^{(m)}=0.
\label{simpl}
\ee
If the discrete derivative operator $\Delta$ is defined by
\be
\Delta \phi_n=\phi_{n+1}-\phi_n
\label{Delta}
\ee
and the shift operator $\hat{E}$ is defined by 
\be
\hat{E}\phi_n\equiv \phi_{n+1},
\label{shift}
\ee
then the homogeneous difference equation (\ref{simpl}) can be rewritten as
\be
\left [\hat{D} \right ]\tilde{\phi}_n^{(m)}\equiv
\left [n\Delta^2 \hat{E}^{-1}+(m+1)\Delta+\lambda\right ]\tilde{\phi}_n^{(m)}=0.
\label{D}
\ee
The difference operator $\left [\hat{D}\right ] $ 
annihilates the field $\tilde{\phi}_n^{(m)} $, $\forall n\in \NN$.

{\bf First solution}

Define for $\alpha$ real, the falling factorial power 
$n^{\frac{\alpha}{ }}$,
\be
n^{\frac{\alpha}{ }}\equiv \frac{\Gamma(n+1)}{\Gamma(n+1-\alpha)},
\qquad 
\stackrel{(\ref{Delta})}{\Longrightarrow}
\qquad
\Delta n^{\frac{\alpha}{ }}=\alpha n^{\frac{\alpha-1}{ }}.
\label{descending}
\ee
For natural $\alpha=k$, $n^{\frac{k}{ }}=n(n-1)\cdots(n-k+1)$;
this explains the name.

We search for a solution of the form
\be
\tilde{\phi}_n^{(m)}=
a^{m}_0(\sigma, \lambda)n^{\frac{\sigma}{ }}+a^{m}_1(\sigma, \lambda)n^{\frac{\sigma+1}{ }}
+a^{m}_2(\sigma, \lambda)n^{\frac{\sigma+2}{ }}+\dots,
\label{expansion}
\ee
i.e. an expansion  in falling factorial powers of $n$.
Equating to zero the coefficient of each $n^{\frac{k+\sigma}{ }}$ in 
\be
\left [\hat{D} \right ] \sum_{k=0}^{\infty} a^{m}_k(\sigma, \lambda) n^{\frac{\sigma+k}{ }}  = 0
\label{tosolve_h}
\ee
we obtain 
the indicial equation
\be
\sigma(\sigma+m)=0  \label{indicial}
\ee
and the recurrence relation for the expansion coefficients $a^{m}_k(\sigma,\lambda)$
\be
(k+\sigma)(k+\sigma+m)\;a^{m}_k(\sigma, \lambda)+\lambda\; a^{m}_{k-1}(\sigma, \lambda)=0. \label{step}
\ee
Eq. (\ref{step}) guarantees that (\ref{expansion}) is also an expansion in powers of $\lambda$.
For $\sigma\rightarrow 0$ we obtain the first finite series solution
\be
u_n^{(m)}=
\sum_{k=0}^{n}\frac{(-\lambda)^k }{k!(m+k)! } 
\left [\frac{\Gamma(n+1)}{\Gamma(n+1-k)}=\frac{n!}{(n-k)!}=n^{\frac{k}{ }}\right ].
\label{u_n^m_h}
\ee
From now on we drop the dimensionfull multiplicative constant $a_0$ by putting $a_0\equiv 1$. 
It can be reintroduced whenever required. 

{\bf Second solution - infinite series}

The second solution cannot be obtained easily since
we confront the case of roots differing by an integer in (\ref{indicial}); 
taking $\sigma=-m$ makes the coefficients diverge after some $k$. 
We therefore use the following procedure \cite{MT}:
we solve the inhomogeneous equation
\be
\left [\hat{D} \right ]
 \sum_{k=0}^{\infty}a_k(\sigma)n^{\frac{\sigma+k}{ }} = 
 \sigma(\sigma+m)^2 n^{\frac{\sigma}{ }} ,
\ee
take the derivative of its solution with respect to $\sigma$, 
and then take the limit $\sigma+m\equiv \epsilon \rightarrow 0$.
This amounts to the evaluation of 
\be
\frac{\partial}{\partial\sigma} 
\left .     \left [
\sum_{k=0}^{\infty} 
\frac
{(-\lambda)^k(\sigma+m)\Gamma(1+\sigma)\Gamma(m+1+\sigma)\Gamma(n+1-k-\sigma)}
{\quad \Gamma(k+1+\sigma) \quad \Gamma(k+m+1+\sigma) \;\; \Gamma(n+1-k-\sigma)} 
\right ]   \right|_{\sigma=-m} \!\!\! .\label{2nd}
\ee
Applying  
$\partial_{\sigma}$ to  
the summand in (\ref{2nd}) one obtains terms involving the digamma function
$\Psi(x)=\frac{d \log \Gamma(x)}{dx}=\frac{\Gamma'(x)}{\Gamma(x)}$. 
Using the poles-displaying expansion (\ref{digamma}), then 
carefully observing the multiplicative cancellation between zeroes and poles 
in the limit $\sigma+m\rightarrow 0$
[and multiplying by a factor $(-)\cdot(m-1)!$ for convenience] 
we obtain the second solution
\bea
w_n^{(m)} &=&
\sum_{k=0}^{n}  \frac{\lambda^m(-\lambda)^k}{(k+m)!}C_n^k(H_{n-k}-H_{k}-H_{k+m}+H_{m-1}-\gamma)
\label{wnm}  \\
&& -  n! \sum_{k=0}^{m-1} \frac{\lambda^k}{k!}\frac{(m-k-1)!}{(m-k+n)!}
+\sum_{k=n+1}^{\infty} \frac{\lambda^{k+m}(-)^{n}n!(k-n-1)!}{k!(k+m)!}.
\nonumber
\eea
As already mentioned, $H_k$ is a discrete version of the logarithmic function,
\be
H_k=1+\frac{1}{2}+\frac{1}{3}+\dots +\frac{1}{k}\; , \quad \; \; k=1,2,3\dots \;\;  ;
\; \; \qquad H_0=0; \label{H_k}
\ee
whereas $\gamma$ is the Euler-Mascheroni constant, 
$\gamma=\lim_{k=\infty}(H_k-ln k) \simeq 0.5772$.

{\bf Second solution - finite series}

Although $w_n^{(m)}$ solves (\ref{simpl}), it is an infinite convergent series in $\lambda$. 
To obtain a finite series,
we search for a linear combination of $u_n^{(m)}$ and $w_n^{(m)}$,
\be
v_n^{(m)}=a(\lambda)u_n^{(m)}+b(\lambda)w_n^{(m)}, 
\label{v_def}
\ee 
that is independent of $\lambda$ in $n=0$ and $n=1$. 
If we choose to impose 
\be
v_0^{(m)}=a(\lambda)u_0^{(m)}+b(\lambda)w_0^{(m)}=0,
\label{v00}
\ee 
we get 
\be
\frac{a(\lambda)}{b(\lambda)}=\; m! \;
\left [ 
\sum_{k=0}^{m-1} \frac{\lambda^k}{k!}\frac{1}{(m-k)}+\frac{\lambda^m}{m!}
\left (\frac{1}{m}+\gamma \right )
-\sum_{k=1}^{\infty}\frac{\lambda^k}{k}\frac{\lambda^m}{(k+m)!}
\right ] .
\label{avsb}
\ee
Using (\ref{wnm}), (\ref{v_def}) and (\ref{avsb}) to evaluate $v_1^{(m)}$ we obtain,
after a calculation featuring unexpected but essential simplifications,
\be
v_1^{(m)}=a(\lambda)u_1^{(m)}+b(\lambda)w_1^{(m)}=\frac{b(\lambda)\;e^{\lambda}}{m+1}.
\ee
Choosing $ b(\lambda)=e^{-\lambda}$  ensures $v_1^{(m)}=\frac{1}{m+1}$, 
which is constant (and convenient).
In consequence,
\be
v_n^{(m)}=
e^{-\lambda }  
w_n^{(m)} +  e^{-\lambda }
\left [
\sum_{k=0}^{m-1} \frac{\lambda^k \; \; m!\;}{k!(m-k)}
+\lambda^m\left (\frac{1}{m}+\gamma\right )
-\sum_{k=1}^{\infty}\frac{\lambda^{k+m}\; m!}{(k+m)!k}
\right ]
u_n^{(m)}  
.
\label{vuw_m}
\ee
We must still substitute in (\ref{vuw_m}) the expression (\ref{u_n^m_h}) for $u_n^{(m)}$ 
and the lenghty expression (\ref{wnm}) for $ w_n^{(m)}$.
This leads to a long calculation involving products and sums of finite and infinite power series.
Making use of the identities 
(\ref{k}-\ref{splendid}), (\ref{plusp1'}-\ref{complicat}) and (\ref{B3})
we obtain the intermediate result
\be
 v_n^{(m)} =  
\sum_{N=0}^{m-1}\lambda^N(m-N-1)!\sum_{k=0}^{n}\frac{(-\lambda)^k C_n^k}{(k+m)!}
- \sum_{N=0}^{m-1}\lambda^N\frac{(m-N-1)!(n+N)!}{N!(m+n)!} +
\label{first_v}
\ee
$$
 + \sum_{N=0}^{n} \frac{(-\lambda)^N\lambda^m}{(N+m)!} 
\left \{
    \sum_{k=0}^{N}C_{N+m}^{k+m}C_n^k(H_{N-k}+H_{n-k}-H_k-H_{k+m})
   +\sum_{j=1}^{m}\frac{C_{m+N-j}^{N}}{n+j}
\right \} . 
$$
At this point, we are still left with a double sum in (\ref{first_v}). 
Two cases appear:

\noindent
1. $n\leq m-1\; . \;$  Then, 
the first term in (\ref{first_v}) can be written as
\be
\sum_{N=0}^{m-1} \lambda^N(m-N-1)! \sum_{k=0}^{n}\frac{(-\lambda)^k C_n^k}{(k+m)!} 
\quad \stackrel{n\leq m-1}{=}  \qquad   \qquad\qquad\qquad
\label{n<m}
\ee
$$  
\left [
\sum_{L=0}^{n-1} \sum_{k=0}^{L}+
\sum_{L=n}^{m-1} \sum_{k=0}^{n}+
\sum_{L=m}^{n+m-1} \sum_{k=L-m+1}^{n}
\right ] 
\left \{
\lambda^L
(-)^kC_n^k\frac{(m+k-L-1)!}{(m+k)!}
\right \} ;
$$

\noindent
2. $n\geq m-1 \;  . \;  $ Now, the same term in (\ref{first_v}) reads
\be
\sum_{N=0}^{m-1}  \lambda^N(m-N-1)!  \sum_{k=0}^{n}\frac{(-\lambda)^k C_n^k}{(k+m)!} 
\quad \stackrel{n\geq m-1}{=}  \qquad   \qquad\qquad\qquad
\label{m<n}
\ee
$$
\left [
\sum_{L=0}^{m-1} \sum_{k=0}^{L}+
\sum_{L=m}^{n-1} \sum_{k=L-m+1}^{L}+
\sum_{L=n}^{n+m-1} \sum_{k=L-m+1}^{n}
\right ] 
\left \{
\lambda^L(-)^kC_n^k\frac{(m+k-L-1)!}{(m+k)!}
\right \} .
$$
Using either (\ref{n<m}) or (\ref{m<n}) - according to the case considered,
grouping terms according to their power of $\lambda$ and
observing some cancellations granted by (\ref{splendid}, \ref{B1}, \ref{B2}),
we arrive both in case (\ref{n<m}) and in case (\ref{m<n}) 
[hence irrespective of the relation between $n$ and $m$], 
at
\be
v_{n}^{(m)}=\sum_{L=0}^{n-1}(-\lambda)^L 
\left \{
\sum_{s=1}^{n-L}\frac{(-)^{s-1}C_{n}^{s+L}}{(m+s)(m+s+1)\cdots(m+s+L)}
\right \}.
\label{third_v}
\ee
This is our final result for $v_{n}^{(m)}$. 
The first $v_n^{(m)}$'s can be calculated easily. 
As expected, $v_0^{(m)}=0$  
for any natural $m$; 
then
\be
v_1^{(m)}=\frac{1}{(m+1)!/m!},
\qquad
v_2^{(m)}=\frac{(3+m)-\lambda}{(m+2)!/m!},
\label{v12m}
\ee
\be
v_3^{(m)}=\frac{(m^2+6m+11)-(2m+8)\lambda+\lambda^2}{(m+3)!/m!}, \quad \forall m\in \NN.
\label{v3m}
\ee
Concerning the functions $f_1$ and $f_2$ in (\ref{param}),
the continuum limit behaviour of $u_n^{(m)}$ and $v_n^{(m)}$ requires, 
cf. Eqs. (\ref{F1}) and (\ref{F2}),
\be
f_1(\lambda)=\lambda^{\frac{m}{2}}, \qquad f_2(\lambda)=\lambda^{-\frac{m}{2}}.
\label{f1f2}
\ee
Putting together (\ref{param}), (\ref{f1f2}), (\ref{u_n^m_h}) and (\ref{third_v}) we obtain our
final result for the two finite series that solve (\ref{Phinm}):
\be
\phi_n^{1(m)}=\sqrt{\lambda^{m}\frac{(n+m)!}{n!}} \sum_{k=0}^{n}
(-\lambda)^k 
\left [
\frac{C_n^k}{(k+m)!}
\right ];
\label{1}
\ee
\be
\phi_n^{2(m)}=\sqrt{\lambda^{-m}\frac{(n+m)!}{n!}}
\sum_{L=0}^{n-1}(-\lambda)^L 
\left \{
\sum_{s=1}^{n-L}\frac{(-)^{s-1}C_{n}^{s+L}(m+s-1)!}{(m+s+L)!}
\right \}.
\label{2}
\ee
The solutions (\ref{1},\ref{2}) are {\em finite sums}
and are linearly independent, since their Casoratian
(the discrete analogue of the Wronskian) is nonvanishing:
\be
D^m_n\equiv\phi_n^{1(m)} \phi_{n+1}^{2(m)}-\phi_{n+1}^{1(m)}\phi_n^{2(m)}=
\frac{1}{\sqrt{n+m+1}\sqrt{n+1}}.
\label{Dmn}
\ee
The general solution of (\ref{Phinm}) is  a linear combination 
$\phi_n^{(m)}=c_1 \phi_n^{1(m)}+c_2\phi_n^{2(m)}$, 
with coefficients $c_1$ and $c_2$ determined from appropriate boundary conditions.
Since  $\phi_n^{(-m)}$ obeys the same equation (\ref{Phinm}), its general form
will also be a superposition of the solutions (\ref{1}) and (\ref{2}) but with
coefficients determined from independently assigned boundary conditions,
$\phi_n^{(-m)}=c'_1 \phi_n^{1(m)}+c'_2\phi_n^{2(m)}$.

Introducing now the expressions found for $\phi_n^{(m)}$ and $\phi_n^{(-m)}$ in (\ref{opformmm})
produces the general solution of the difference equation of motion (\ref{Phinm}).
As already proved in Section 3,
$m$ (respectively $-m$) characterizes simultaneously the degree of nonlocality 
and the angular momentum of each bilocal configuration 
$\ket{n+m}\phi_n^{(m)}\bra{n}$ (respectively $\ket{n}\phi_n^{(-m)}\bra{n+m}$) in (\ref{opformmm}). 
The coefficients $a_m$ and $b_m$ are determined by the imposed initial/boundary conditions,
for instance $a_m=\delta_{m,m_0}$ and $b_m\equiv 0$
if only a field configuration of angular momentum $m\omega$ is excited.

{\bf Connection to Laguerre polynomials}

It is worthwhile to mention a simple mathematical connection.
If one writes $\phi_n^{(m)}\sim\sqrt{\frac{n!}{(n+m)!}}L_n^{(m)}$
[in contradistinction to (\ref{subst})] one obtains
\be
(n+1)L_{n+1}^{(m)}+(n+m)L_{n-1}^{(m)}+(\lambda-m-2n-1)L_n^{(m)}=0.
\label{Lrr}
\ee
This is precisely the recurrence relation for the Laguerre polynomials,
and it has {\em two} independent solutions expressible as polynomials in $\lambda$.
The first, $L_n^{1(m)}=\frac{(n+m)!}{n!}u_n^{(m)}$, 
gives the well-known Laguerre polynomials (\ref{L_n^m}).
The second finite series solution of (\ref{Lrr}) provides 
some kind of  "Laguerre polynomials of the second kind" $L_n^{2(m)}$
[which obey the Laguerre recurrence relation in $n$, (\ref{Lrr}), 
 but not the Laguerre differential equation in $\lambda$]
related to our $ v_n^{(m)}$ through $L_n^{2(m)}=\frac{(n+m)!}{n!}v_n^{(m)}$. 
Hence, we also obtained in this subsection a second finite series solution
of the Laguerre recurrence relation (\ref{Lrr}), apparently unavailable in the literature.

{\bf Radially symmetric solutions}

If one has radial symmetry, $\phi=\phi(\hat{N})$, the relevant difference equation simplifies
to the local form (\ref{em3}).
Its first solution \cite{A1} follows immediately by taking $m=0$ in (\ref{u_n^m_h})
\be
u_n\equiv u^{(0)}_n=
\sum_{k=0}^{n} \frac{(-\lambda)^k}{k!}\;  C^k_n \; , \label{phi1}
\ee
To obtain the second solution one is confronted when $m=0$  with an indicial equation $\sigma^2=0$
that has roots which are equal, not differing by an integer. In spite of slight differences
in proceeding due to that aspect, the resulting infinite series solution 
\be
w_n 
= \sum_{k=0}^{n}\frac{(-\lambda)^k}{k!}
C_n^k (H_{n-k}-2H_k-\gamma)
+\frac{(-\lambda)^n}{n!}\sum_{s=1}^{\infty}\frac{(+\lambda)^{s}(s-1)!}{[(n+s)!/n!]^2}
\label{wn}
\ee
is still the specialization of (\ref{wnm}) to $m=0$, namely $w_n\equiv w_n^{(0)}$.
The last term in the right hand side of (\ref{wn}) was missed in \cite{A1}.
Being negligible in the commutative limit, 
this term does not affect the large distance behaviour.

Searching for a polynomial solution $v_n=a(\lambda)u_n+b(\lambda)w_n$ that obeys
$v_0=0$ and $v_1=1$  
one  obtains  the $m=0$ specialization 
of (\ref{first_v}),
\be
v_n=\sum_{L=0}^{n-1}\frac{(-\lambda)^L}{L!}\sum_{k=0}^{L}C_n^kC_L^k(H_{L-k}+H_{n-k}-2H_k) . 
\label{CCH} 
\ee
The form (\ref{CCH}), which appeared first in \cite{cfw} without proof,
can be considerably simplified.
Namely, using (\ref{egal1}) one obtains
\be
v_n=\sum_{N=0}^{n-1}(-\lambda)^N
\left \{
\sum_{s=1}^{n-N}\frac{(-)^{s-1}C_n^{s+N}}{s(s+1)\cdots(s+N)}
\right \}
\label{v3}
\ee
which could have been obtained directly by taking $m=0$ in our general result (\ref{third_v}) .
Further use of (\ref{splendid}) allows to reach an even simpler expression
\be
v_n=\sum_{k=0}^{n-1} \frac{(-\lambda)^k}{k!} \;\sum_{j=1}^{n-k}\frac{C^k_{n-j}}{k+j}\; . 
\label{phi2}
\ee
Both forms (\ref{v3}) and (\ref{phi2}) are useful: 
the first  is handier for some analytical manipulations
whereas the second  permits faster numerical evaluations.

\section{Sources}

Sources can be taken into account by introducing an inhomogeneous term $j_n$ in (\ref{Phinm}),
\be
\sqrt{n+m+1}\sqrt{n+1}\Phi^{(m)}_{n+1}+\sqrt{n+m} \sqrt{n} \Phi^{(m)}_{n-1}
+(\lambda-2n-m-1) \Phi^{(m)}_{n}=j_n.
\label{Phinm_j}
\ee
The solution of Eq. (\ref{Phinm_j}) takes into account an arbitrary distribution of sources $j_n$
and is consequently denoted by $ \Phi^{(m)}_{n}[j_n]$. 
It can be obtained through the linear superposition of solutions 
$ \Phi^{(m)}_{n}[j\delta_{n,n_0}]$ which solve Eq. (\ref{Phinm_j}) with  sources $j_n=j\delta_{n,n_0}$
localized at an arbitrary but single point $n_0$. 

If $n_0=0$ one has the most interesting case - a source at the origin, $j_n=j\delta_{n,0}$.
If  $j=\sqrt{m!}\lambda^{-m/2}$,
the difference equation (\ref{Phinm_j}) is now solved precisely by 
 $ \phi_n^{2(m)}$ ($+c\phi_n^{1(m)}$, $c$ arbitrary).
 Indeed, our hardly won $\phi_n^{2(m)}$
does solve the homogeneous equation  (\ref{Phinm}) everywhere except at $n=0$, simply
due to the fact that there the difference equation becomes first order and admits
only one solution, which is $\phi_n^{1(m)}$. 
If a source $ j=\sqrt{m!}\lambda^{-m/2}$
is added at $n=0$ however, $\phi_n^{2(m)} $ is the particular solution of the 
resulting inhomogeneous equation.
This is in line with the fact that $v_n$ enters the description of radially propagating waves, 
which {\em require} a source at the origin.
If we take $j$ arbitrary in $j_n=j\delta_{n,0}$, 
the solution will be (we write $j\equiv j_0$ for later convenience)
\be
\Phi^{(m)}_{n}[j_0\delta_{n,0}]=j_0\frac{\lambda^{m/2}}{\sqrt{m!}}\; \phi_n^{2(m)}.
\label{source_at_origin}
\ee 
 
Consider then the case of a source at $n_0\geq 1$ (a ring-like source). 
We search for a solution of (\ref{Phinm_j}) with source $j_n=j\delta_{n,n_0}$. 
A straightforward adaptation to the discrete case of the method of variation of constants 
(or of the method of Green functions) suggests to search for 
\be
\Phi_n^{(m)}[j\delta_{n,n_0}]=c_1(n)\phi_n^{1(m)} + c_2(n)\phi_n^{2(m)},
\ee
with $c_1$ and $c_2$ jumping only between $n_0$ and $n_0+1$, namely 
\be
c_1(n+1)-c_1(n)=d_1\delta_{n_0,n},  \qquad c_2(n+1)-c_2(n)=d_2\delta_{n_0,n}.
\ee
Eqs. (\ref{Phinm_j}) and (\ref{Phinm}) imply then
\be
d_1=\frac{-j\;\phi_{n_0}^{2(m)}}{\sqrt{n_0+m+1}\sqrt{n_0+1}D^{m}_{n_0}}, \qquad 
d_2=\frac{+j\;\phi_{n_0}^{1(m)}}{\sqrt{n_0+m+1}\sqrt{n_0+1}D^{m}_{n_0}},
\ee
where $D^{m}_{n_0}$ is the Casoration of $\phi_n^{1(m)}$ and $\phi_n^{2(m)} $ at $n=n_0$, 
cf. Eq. (\ref{Dmn}).
The solution for a source $j\delta_{n,n_0}$ of intensity $j$ and location $n_0$
is consequently
\be
\Phi_n^{(m)}[j\delta_{n,n_0}]=c_1 \phi_n^{1(m)} + c_2 \phi_n^{2(m)}
-j \theta_{n,n_0}
\left (\phi_n^{1(m)}\phi_{n_0}^{2(m)}-\phi_n^{2(m)}\phi_{n_0}^{1(m)}
\right ),
\ee
with the step function $\theta_{m,n}\equiv \theta (m-n)$ defined as
\be
\theta_{m,n}=  \left \{
\begin{array}{ll}
1 & \mbox{if  } m-n\geq 0 \\
0 & \mbox{if  } m-n<0.  
\end{array}    \right .
\ee
Finally, denoting by 
$\Phi_n^{(m)}[0]=c_1 \phi_n^{1(m)} + c_2 \phi_n^{2(m)}$ 
the general solution in the absence of sources,
we obtain
the general solution with an arbitrary source distribution $j_n$ as
\be
\Phi_n^{(m)}[j_n]=\Phi_n^{(m)}[0]-\sum_{n_0}j_{n_0}\theta_{n,n_0}
\left (\phi_n^{1(m)}\phi_{n_0}^{2(m)}-\phi_n^{2(m)}\phi_{n_0}^{1(m)}
\right ) .
\label{arbitrary_j}
\ee
The sum is taken over all points $n_0$ with nonzero sources, $j_{n_0}\neq 0$.
If a source $j_0$ appears at the origin one notes that,
due to $\phi_{0}^{1(m)}=\frac{\lambda^{m/2}}{m!} $ and $ \phi_{0}^{2(m)}=0$,
the $n_0=0$ contribution in (\ref{arbitrary_j}) reproduces (\ref{source_at_origin}).
The general solution (\ref{arbitrary_j}) does not display singularities, 
even at the location of the sources.

\section{Commutative limit : types of waves}

It is useful to see first what happens to the difference equation (\ref{Phinm}) in the commutative limit.
As $\theta\rightarrow 0$, one has $n\simeq n'\simeq \frac{r^2}{2\theta}\rightarrow\infty$
and $\lambda=\frac{\theta\omega^2}{2}\rightarrow 0$,
but $\lambda\cdot n\sim (\frac{\omega r}{2})^2$ is finite;
$m$ stays finite as well.
In this limit the difference operator applied in (\ref{Phinm}) to a function of $n$, $\phi^{(m)}_{n}$, 
becomes the $m$-th order Bessel operator applied to a function of $r$, call it $f^{(m)}(r)$.
Indeed, taking $ n\rightarrow \infty$, 
expanding the square roots in (\ref{Phinm}) to order $O(\frac{1}{n^2})$
and replacing $\frac{\Delta\;\;}{\Delta n}$ by $\frac{d\;\;}{dn}$ one obtains
\be
\left (n+\frac{m}{2}\right ) \frac{d^2\phi_n^{m}}{dn^2} +
\frac{d\phi_n^{m}}{dn} +
\left ( \lambda- \frac{m^2}{4n} \right )\phi_n^{m}=0.
\label{ninfinite}
\ee 
Recalling that $\lambda=\frac{\theta\omega^2}{2}$ 
and passing via $n\equiv\frac{r^2}{2\theta}$ from a function of $n$ to a function of $r$,
$\phi^{(m)}_{n} \rightarrow f^{(m)}(r)$, Eq. (\ref{ninfinite}) becomes as $\theta\rightarrow 0$:
\be
\frac{d^2f^{(m)}}{dr^2}+\frac{1}{r}\frac{df^{(m)}}{dr}+(\omega^2-\frac{m^2}{r^2})f^{(m)}(r)=0.
\ee
This is precisely the Bessel equation of order $m$ for a function of independent variable $\omega r$;
The solutions of (\ref{Phinm}) should therefore reduce at large distances 
to linear combinations of the cylindrical functions of order $m$, (\ref{Jm}) and (\ref{Ym}).

Our solutions are consistent with the above limit. Indeed,
as $n\simeq\frac{r^2}{2\theta}\rightarrow \infty$, 
$\lambda=\frac{\theta\omega^2}{2}\rightarrow 0$ and 
$ \lambda\cdot n \simeq \left (\frac{\omega r}{2}\right )^2$, 
Eqs. (\ref{u_n^m_h}) and (\ref{Jm}) imply that a properly normalized $ u_n^{(m)}$
becomes, as a function of $r$, the $m$-th order Bessel function $J_m(\omega r)$,
\be
\sqrt{\lambda^{m}\frac{(n+m)!}{n!}}\; u_n^{(m)} \longrightarrow J_m(\omega r).
\label{F1}
\ee
In this way we also establish the function $f_1$ in (\ref{param}) to be $\lambda^{m/2}$,
up to multiplicative factors which go to 1 when $\lambda$ goes to 0.
For instance, ortogonality of the $\phi_n^{1,m}$ in (\ref{1}), seen as functions of  $\lambda\in[0,\infty)$,
requires further multiplication of the LHS of (\ref{F1}) by $\e^{-\frac{\lambda}{2}}$. 
Not being of immediate relevance for our purposes, such factors are omitted.

For the second solution things are less immediate. 
It is convenient to first consider $ w_n^{(m)}$. When $n\rightarrow\infty$ and
$\lambda\rightarrow 0$ various terms disappear. First, the last one in (\ref{wnm}) vanishes,
\be
\left |\sum_{s=1}^{\infty}\frac{(-\lambda)^n\lambda^s (s-1)! n!}{(s+n)!(s+n+m)!} \right | \leq
\sum_{s=1}^{\infty}\frac{\lambda^{n+s}}{(s+n)!}\frac{1}{n+s} \leq \frac{e^{\lambda}}{n}\rightarrow 0.
\label{sumaS}
\ee
Second, $v_n^{(m)} $ an $w_n^{(m)} $ have the same commutative limit,
due to the fact that the ratio $\frac{a}{b}$ in (\ref{avsb}) vanishes as $\lambda\rightarrow 0$.
Using this observation and Eqs. (\ref{wnm}), (\ref{Jm}) and (\ref{Ym}) we obtain
\be
\sqrt{\lambda^{-m}\frac{(n+m)!}{n!}}v_n^{(m)} \longrightarrow \pi Y_m(\omega r)
+J_m(\omega r) [H_{m-1}-2\gamma -\log (\theta\omega^2/2)].
\label{F2}
\ee
This also establishes the function $f_2$ in (\ref{param}) to be $\lambda^{-m/2}$.

The correspondence between NC and usual waves can now be established through their behaviour
at large distances. 
Given that 
the $m$-th order Bessel function $J_m(\omega r)$ describes usual radially standing waves (oscillations),
Eq. (\ref{F1}) implies that its counterpart
$\phi_n^{1(m)}$ describes radially standing NC waves (oscillations).
Moreover, $J_m(\omega r)$ and $Y_m(\omega r)$ can carry
angular momentum $\omega m$ or $-\omega m$ in usual planar field theory, 
in perfect agrement with the fact that $ \phi^{1(m)}_n $ enters (\ref{opformmm}) 
in combination with either $\ket{n+m}\bra{n}$ or $\ket{n}\bra{n+m}$. 

On the other hand, the first Hankel function of order $m$ 
\be
H^1_m(\omega r)=J_m(\omega r)+iY_m(\omega r)
\ee
describes waves which propagate outward radially 
and rotate angularly with frequency plus or minus $\omega m$ 
[unless waves with $m$ and $-m$ dependence are superposed, 
e.g. to render the angular part of the wave standing].
In consequence, the linear combination of $ \phi^{1(m)}_n$ and $\phi^{2(m)}_n $ 
which tends to $ H^1_m(\omega r)$ as $\theta\rightarrow 0$ will describe
a NC wave radially propagating outwards towards $n=\infty$ 
and carrying angular momentum $+m \omega$ or $-m \omega$ 
[unless two waves with opposite $m$ are superposed]. 
This combination is easily found to be
\be
\phi^{3(m)}_n=\phi^{1(m)}_n+\frac{i}{\pi}
\left (
\phi^{2(m)}_n-\phi^{1(m)}_n 
\left [ 
H_{m-1}-2\gamma-\log \frac{\theta\omega^2}{2}
\right ]
\right )
\ee
and displays angular momentum $m\omega$ when it combines with $\ket{n+m}\bra{n}$
and angular momentum $-m\omega$ when it combines with $\ket{n}\bra{n+m}$.
Similarly, the easily found linear combination of $ \phi^{1(m)}_n$ and $\phi^{2(m)}_n $ which,
as $ \theta \rightarrow 0$, tends to 
$ H^2_m(\omega r)=J_m(\omega r)-iY_m(\omega r)$ 
will describe "radially collapsing" NC waves.

\section{Appendix - Useful Formulae}

We collect here various formulae useful for the mathematical manipulations in the paper.
To date, Eqs. 
(\ref{k[n]}), (\ref{splendid}), (\ref{plusp---'}), (\ref{egal1}), (\ref{B1}), (\ref{B3})  
could not be found in the literature accessed by the author.
The form of these [possibly new] identities was suggested by the problem studied. 
Their proof requires only elementary methods and, occasionally, some effort.

Recall first that $H_n\equiv 1+\frac{1}{2}+\frac{1}{3}+\dots+\frac{1}{n}$. 
Then we have the known identity
\be
\sum_{k=1}^{n}\frac{(-)^{k-1}C_n^k}{k}=H_n
\label{k}
\ee
together with two useful generalizations of it,
\be
\sum_{k=1}^{n}\frac{(-)^{k-1}}{k}\frac{C_n^k \quad m!}{(k+1)(k+2)\dots(k+m)}=H_{n+m}-H_m,
\label{k[n]}
\ee
\be
\sum_{k=1}^{n}\frac{(-)^{k-1}}{k}\frac{C_{n+m}^{k+m}\quad p!}{(k+1)\dots(k+p)}=
\sum_{j=1}^{n}\frac{C^m_{m+n-j}}{p+j}.
\label{splendid}
\ee
One can translate the denominator $k$ in (\ref{k}) by a positive integer amount $p$, 
to find the identity (this time one can extend the sum to $k=0$)
\be
\sum_{k=0}^{n}\frac{(-)^{k-1}C_n^k}{k+p}=-\frac{(p-1)!n!}{(p+n)!} 
\qquad p=1,2,3\dots (p\geq 1).
\label{plusp1'}
\ee
Its generalization is, again for $p$ integer and  $p\geq 1$,
\be
\sum_{k=0}^{n}\frac{(-)^{k-1}C_n^k\; m!}{(k+p)(k+p+1)\dots(k+p+m)}
=-\frac{(p-1)!(n+m)!}{(p+n+m)!}.
\label{plusp---'}
\ee
The use of 
$C_{n+m}^{k+m}=\sum_{j=0}^{n-k}C_{m-1+j}^{m-1}C_{n-j}^{k}$ allows for a further extension:
\be
\sum_{s=1}^{n}\frac{(-)^{s-1}C_{n+L}^{s+L}}{(s+p)\cdots(s+p+m)}
=\sum_{j=0}^{n-1}C_{L-1+j}^{L-1}
\sum_{s=1}^{n-j}\frac{(-)^{s-1}C_{n-j}^{s}}{(s+p)\cdots(s+p+m)},
\label{complicat}
\ee
after which one can use (\ref{plusp---'}) to complete the evaluation of the sum.

\bigskip

The digamma function $\Psi(x)=\frac{d \log \Gamma(x)}{dx}=\frac{\Gamma'(x)}{\Gamma(x)}$,
has the following useful properties \cite{A-S}
($z$ complex but not a negative integer, $n$ positive integer) 
\be
\Psi(z+1)\equiv \frac{\Gamma'(z+1)}{\Gamma(z+1)}=-\gamma +\sum_{l=1}^{\infty}\frac{z}{l(l+z)},
\quad
\Psi(n+1)=-\gamma +H_n.
\label{digamma}
\ee
The series expansion of the Bessel functions \cite{A-S}
is required while taking the commutative limit of $u_{n}^{(m)}$,
\be
J_m(z)=\sum_{k=0}^{\infty}\frac{(-)^k(z/2)^{2k+m}}{k!(k+m)!},
\label{Jm}
\ee
while the Neumann functions $Y_m$ \cite{A-S} are also needed for $w_{n}^{(m)}$,
\bea
\pi Y_m(z)&=& 2J_m(z)(\gamma+\log\frac{z}{2})-
\sum_{k=0}^{m-1}\frac{(m-k-1)!}{k!}\left ( \frac{z}{2}\right )^{2k-m} \nonumber  \\ 
&&-\sum_{k=0}^{\infty}(-)^k \left ( \frac{z}{2}\right )^{2k+m} \frac{H_{m+k}+H_k}{k!(k+m)!}.
\label{Ym}
\eea
The Laguerre polynomials are given by \cite{A-S}
\be
L_n^m(\lambda)=\sum_{k}^{n}\frac{\Gamma(n+m+1)}{\Gamma(k+m+1)}\frac{(-\lambda)^k}{k!(n-k)!}.
\label{L_n^m}
\ee

\bigskip

Finally, we write down
the identities ensuring that $v_n^{(m)}$ is a polynomial of order $n-1$ in $\lambda$.
If $m=0$ the following suffices, for $N\geq n+1$:
\be
\sum_{k=0}^{n}C_N^kC_n^k(H_{N-k}+H_{n-k}-2H_k) 
=\sum_{s=1}^{N-n}\frac{(-)^{s-1}
\left \{
C_{N}^{s+n}=C_{(N-n)+n}^{\; \; \; \; \;s \;\; \; \; +n}
\right \}
n!}{s\cdot(s+1)\cdots(s+n)}  
\label{egal1} 
\ee
For $m\geq 1$ however, we have two cases.
If $N\leq n $, introduce for convenience the notation
\be
\sum_{k=0}^{N}C_{N+m}^{k+m}C_n^k(H_{N-k}+H_{n-k}-H_k-H_{k+m})+\sum_{j=1}^{m}\frac{C_{m+N-j}^{N}}{n+j}
\equiv
A_{n,N}^{m}.
\label{notatie}
\ee
Then
\be
\frac{(-)^{N}}{(N+m)!} A_{n,N}^{m}=
\sum_{k=N+1}^{n}(-)^{k-1}C_n^k\frac{(k-N-1)!}{(m+k)!} 
\label{B1}
\ee
and
\be
A_{n,n}^{m}\equiv 0.
\label{B2}
\ee
On the other hand, if $ N\geq n+1$ the relevant identity is 
\be
\sum_{k=0}^{n}C_{N+m}^{k+m}C_n^k(H_{N-k}+H_{n-k}-H_k-H_{k+m})
+\sum_{j=1}^{m}\frac{C_{m+N-j}^{N}}{n+j}
=
\sum_{j=1}^{N-n}\frac{C_{m+N-j}^{n+m}}{n+j}.
\label{B3}
\ee


\subsection*{Acknowledgements}

We acknowledge financial support from the CNCS through 
the  {\it Idei}  Program Contract Nr. 121/2011 and 
the {\it Nucleu} Program Contract PN-09370102/2009.

\end{document}